\documentclass[10pt]{article}
\usepackage{calc}
\usepackage{a4wide}
\usepackage[ngerman,english]{babel}
\usepackage{etex}
\usepackage[T1]{fontenc}
\usepackage{ellipsis,ragged2e}
\usepackage{microtype}
\usepackage{amsmath}
\usepackage{amssymb}
\usepackage{amsthm}
\usepackage{color}
\usepackage{times}
\usepackage{pictexwd}
\usepackage[pdftex]{graphicx}
\usepackage{graphicx}
\graphicspath{{figures/}}
\usepackage{caption}
\usepackage{subcaption}
\usepackage{hyperref}
\usepackage{cite}
\usepackage{mathrsfs,graphicx,color,float, indentfirst,textcomp}
\usepackage{setspace}
\usepackage{latexsym,amssymb,lscape,amsmath,amsthm,amsfonts,amssymb,cite,enumerate}
\usepackage{lscape}
\usepackage{xfrac}
\usepackage{indentfirst}

\usepackage{algorithmic}
\usepackage{algorithm}
\usepackage[subnum]{cases}

\usepackage{geometry}

\newtheorem{mydef}{Definition}
\newtheorem{mytheo}{Theorem}

\newcommand\dv{\Delta v}

\renewcommand\d{\mathrm{d}}

\newcommand\Maxw{\mathrm{M}}
\newcommand\HV{\mathsf{H}}
\newcommand\AV{\mathsf{A}}

\newcommand{\fh}{f_\HV}
\newcommand{\fa}{f_\AV}

\newcommand{\rmax}{\rho_{\max}}
\newcommand{\vmax}{v_{\max}}
\newcommand{\wmin}{w_{\min}}

\usepackage[textsize=footnotesize,backgroundcolor=yellow!70,bordercolor=orange]{todonotes}
\usepackage{tcolorbox}
\newcommand{\revision}[1]{\textcolor{black}{#1}}

\title{Model of vehicle interactions with autonomous cars and its properties}
\date{\today}
\author{
	Michael Herty \medskip\\
	{\small\it Institut f\"{u}r Geometrie und Praktische Mathematik (IGPM)} \\
	{\small\it RWTH Aachen University} \\
	{\small\it Templergraben 55, 52062 Aachen, Germany}
	\bigskip \\
	Gabriella Puppo \medskip\\
	{\small\it Dipartimento di Matematica ``Guido Castelnuovo''} \\
	{\small\it ``La Sapienza'' Universit\`{a} di Roma} \\
	{\small\it Piazza Aldo Moro 5, 00185 Roma, Italy}
	\bigskip \\
	Giuseppe Visconti \medskip\\
	{\small\it Dipartimento di Matematica ``Guido Castelnuovo''} \\
	{\small\it ``La Sapienza'' Universit\`{a} di Roma} \\
	{\small\it Piazza Aldo Moro 5, 00185 Roma, Italy}
}

\begin{document}
	
\maketitle

\begin{abstract}
	We study a hierarchy of models based on kinetic equations for the descriptions of traffic flow in presence of autonomous and human--driven vehicles. \revision{The autonomous cars considered in this paper are thought of as vehicles endowed with some degree of autonomous driving which decreases the stochasticity of the drivers' behavior. Compared to the existing literature, we do not model autonomous cars as externally controlled vehicles. We investigate whether this feature is enough to provide a stabilization of traffic instabilities such as stop and go waves. We propose two indicators to quantify traffic instability and we find, with analytical and numerical tools, that traffic instabilities are damped as the penetration rate of the autonomous vehicles increases.}
\end{abstract}



\section{Introduction} \label{sec:introduction}

The analysis, prediction and control of traffic flow are important aspects of the modern approach to vehicular traffic. The increase in the number of circulating vehicles together with the potentialities of the assessment of local conditions \revision{and the possibility of control} prompt the need for better \revision{understanding} of vehicular \revision{traffic and provide} the tools for hopefully achieving improved \revision{flow}. At the same time, the awareness of issues such as safety and sustainability increase the public concern towards the global consequences of traffic~\cite{Chow}.

\revision{Empirical evidence shows} that vehicular traffic is slowed down by instability phenomena, such as stop and go waves, which arise in congested flow. Stop and go waves determine a decrease of the overall flow along highways, thus decreasing the efficiency of our motorways~\cite{Wang,Laval}. \revision{Moreover,} unstable phenomena increase the unpredictability of traffic flow, and thus increase the risk of accidents~\revision{\cite{KuangQuYan,KuangKuWang,KuangEtAl,WangEtAl}. See also} the recent report~\cite{WHO}. Recently, vehicles enriched with a partial degree of automated \revision{features} have been released~\revision{\cite{9046805,Singh_2021}}. Experiments~\cite{SternEtAl2018,DelleMonache2019} have shown that traffic stability can be increased by introducing in the flow a certain number of \emph{externally controlled} vehicles, whose only purpose is  to interact with the human driven vehicles, damping unstable phenomena. On the other hand, it is interesting to \revision{understand} whether vehicles endowed with some degree of \emph{automatic control} have the potentialities to improve traffic stability, as they travel together with human--driven cars. \revision{We stress the difference between externally controlled and automatically controlled or autonomous vehicles. The former behave according to signals sent from outside the vehicle due to external centralized decisions. The latter, instead, are capable of monitoring the environment, provide precise and instantaneous information which influence the human driver, thus decreasing the stochasticity of her/his decisions. Ultimately, one can think that this information feeds the algorithm driving a fully automated car.}

Stop and go waves are \revision{approximately} periodic \revision{in space} disturbances in the density of vehicles~\cite{SeiboldFlynnKasimovRosales2013,Seibold_BookICIAM}, which propagate backwards with respect to the direction of the flow. Mathematically, the classical \revision{macroscopic} description of traffic flow is given by the Lighthill--Whitham--Richards (LWR) model~\cite{lighthill1955PRSL,richards1956OR}, which is based on a single conservation law. Thus, in the LWR model, there is no room for stop and go waves, because the solution of a single conservation law satisfies a maximum principle, which states that no new minima or maxima can arise during the flow. For this reason, several more sophisticated models have been proposed, for instance considering hysteresis to the LWR model~\cite{CorliFao} or modeling traffic by a system of hyperbolic equations\revision{, starting from the well--known works by Aw and Rascle~\cite{aw2000SIAP} and Zhang~\cite{Zhang2002}.}

From a kinetic point of view~\revision{\cite{coscia2007IJNM,DelitalaTosin2007,FermoTosin13,klar1997Enskog,KlarWegener96,PrigogineHerman}}, one can embed the LWR model into a kinetic model, based on microscopic interactions between vehicles~\cite{PgSmTaVg3}. In this view, the LWR model can be obtained as the relaxation equation the kinetic model decays to in its evolution toward \revision{the local equilibrium state}. In kinetic models, the fundamental diagram of the LWR model, i.e. the flux of vehicles as a function of their density, is obtained naturally from the equilibrium distribution of the kinetic model. In this view, stop and go waves are non equilibrium phenomena that arise from the stochastic behavior of individual drivers. 

It is well known that standard kinetic models close to equilibrium \revision{are approximated by} a non linear convection diffusion equation \revision{for the evolution of the density $\rho$}, with a diffusion coefficient $\mu(\rho)$\revision{. In traffic flow $\mu(\rho)$} becomes {\em negative} in the congested regime. Several attempts have been envisaged to modify the models in order to obtain a positive $\mu(\rho)$ which would result into a stable flow \cite{BorscheKlar}. In our works instead~\cite{HPRV20,HPV_Book}, we have recognized the fact that the sign of $\mu$ can signal the onset of unstable waves, which are a known feature of vehicular traffic.

In this work, we consider a mixture of autonomous and human driven vehicles \revision{within the kinetic model introduced in~\cite{HPRV20}. We characterize the interactions between human--driven and autonomous vehicles by assuming that human--driven vehicles adapt their speeds stochastically to the speed of the vehicles ahead, while autonomous cars react deterministically to local conditions. Our purpose is to investigate whether vehicles characterized by a deterministic reaction to the local features of the flow can have a stabilizing influence over overall traffic, by the mere fact of reducing the stochasticity of the flow. We} identify two global parameters of the flow as indices of instability. First, we study the sign of the local viscosity of the flow, $\mu(\rho)$\revision{, derived from a Chapman--Enskog expansion of the kinetic equation}. More precisely, we investigate the amplitude of the interval $(\alpha(p), \beta(p))\in [0,\rmax]$ \revision{of densities $\rho$},
on which $\mu(\rho)$ is negative\revision{. The boundaries of this interval depend on $\rmax$, that is the maximum density of the flow, and on} the percentage of autonomous cars, or penetration rate, $p$. We show that \revision{the interval $(\alpha(p), \beta(p))$ moves towards highest values of the density as $p$ increases and moreover that its amplitude
shrinks as $p$ grows towards 1.}
Secondly, we study the variance of the microscopic speeds of the vehicles, which again decreases, as $p$ grows. This again is a measure of the departure from equilibrium of the flow. We are able to prove our results in a very simplified case, but the overall features of the stabilization effects of autonomous cars are confirmed by numerical solutions in a more general setting.

\revision{Similar evidence is discussed e.g.~in~\cite{Piccoli2020,TosinZanella2021} where autonomous vehicles are externally controlled, i.e.~the authors assume that microscopic interactions are influenced by an external control. Therefore, the interesting point of this work, we believe, is that the stabilization effect of autonomous cars is achieved by their simple presence in the flow, without need to assume that there is an external entity influencing the flow.}

The paper is organized as follows. In Section~\ref{sec:review} we review the kinetic models of~\cite{PSTV2,HPRV20} which represent the starting point for the extension to the mixture of autonomous and human--driven cars in Section~\ref{sec:autonomous}. We \revision{analyze} the properties of the kinetic model \revision{for the single kinetic distribution} and \revision{we discuss} the effects on the traffic stabilization both from an analytical and numerical point of view. We conclude the paper in Section~\ref{sec:conclusion}.

\section{A kinetic model for vehicular traffic} \label{sec:review}

\subsection{Preliminaries on mesoscopic scale models} \label{sec:review:kinetic}

A kinetic model provides a mesoscopic scale representation of microscopic physics described by binary interactions among particles. This is also the case of vehicular traffic flow, where particles are identified by the vehicles on the road. Classical kinetic equations are typically synthesized by the following partial differential equation: 
\begin{equation}\label{eq:generalKineticModel}
	\partial_t f(t,x,v) + v \partial_x f(t,x,v) = \gamma Q[f,f](t,x,v),
\end{equation}
where $f(t,x,v)\, : \mathbb{R}^+_0 \times \mathbb{R} \times \mathcal{V} \rightarrow \mathbb{R}^+_0$ is the mass distribution function and the source term, commonly known as collision kernel, models the change of $f$ due to the change of the microscopic states as a consequence of interactions among particles. The quantity $\gamma>0$ yields a relaxation rate weighting the relative strength between the convective term and the source term.

From the kinetic distribution $f(t,x,v)$ it is possible to recover the local macroscopic density of the particles at time $t$ and space $x$, which we denote by $\rho(t,x)$, as 
\begin{equation} \label{eq:density}
	\rho(t,x) = \int_{\mathcal{V}} f(t,x,v) \d v.
\end{equation}
We always assume that the density is bounded by a maximum value $\rmax\in\mathbb{R}^+_0$. Other aggregate macroscopic quantities of interest are the mean speed and the flux of the particles which are defined in terms of the first moment of $f$ as, respectively,
\begin{equation*} \label{eq:speed_flux}
u(t,x) = \frac{1}{\rho(t,x)} \int_{\mathcal{V}} v f(t,x,v) \d v, \quad q(t,x) = (\rho u)(t,x) = \int_{\mathcal{V}} v f(t,x,v) \d v.
\end{equation*}

The study of kinetic model goes through the analysis of the corresponding spatially homogeneous formulation. For instance, in the Boltzmann--type setting, equation~\eqref{eq:generalKineticModel} writes
\begin{equation}\label{eq:homogBoltzmann}
	\partial_t f(t,v) = \gamma \int_\mathcal{V} \int_\mathcal{V} \mathcal{P}(v_*\to v | v^*;\rho) f(t,v_*) f(t,v^*) \d v_* \d v^* - \gamma \rho f(t,v) = Q[f,f](t,v).
\end{equation}

\revision{In classical kinetic models for gas dynamics, the Boltzmann equation is obtained with two hypotheses. Binary interactions are based on the idea that the flow is rarefied, so that the probability of simultaneous multiple collisions is negligible. Further, the Boltzmann equation is based on the molecular chaos hypothesis which permits to write the probability distributions for the two colliding particles as the product of the single particle distribution functions. In traffic flow, binary interactions are due to the fact that each vehicle reacts only to the vehicle ahead, irrespective of the density of the flow. Moreover, we assume that just before the interaction the velocities of the two interacting vehicles are uncorrelated, which implies that the two particle joint probability can be written as the product of the single particle distributions. This occurs in particular in the space homogeneous case~\eqref{eq:homogBoltzmann}.}

The spatially homogeneous model\revision{~\eqref{eq:homogBoltzmann} is important because it} allows to characterize the stationary state, i.e.~a solution to $Q[f,f](t,v)=0$. In the case there exists one, we denote it by $\Maxw_f(v;\rho)$ and call it Maxwellian distribution. Obviously, the steady state solution depends on the choice of the operator $\mathcal{P}$ that prescribes, in a probabilistic way, the outcome of an interaction between two vehicles, the leading car with velocity $v^*$ and the trailing car with velocity $v_*$. The latter changes then the velocity to $v$, which is the post--interaction velocity defined by the microscopic interaction rules. The core is then the modeling of $\mathcal{P}$ and we discuss in Section~\ref{sec:review:homogeneous} the model presented in~\cite{PSTV2}.

The Maxwellian state allows to define also the mean speed and the flux of vehicles at equilibrium as
\begin{equation} \label{eq:macroQuantitiesEq}
	U(\rho) = \frac{1}{\rho} \int_{\mathcal{V}} v \Maxw_f(v;\rho) \d v, \quad F(\rho) = \rho U(\rho) = \int_{\mathcal{V}} v \Maxw_f(v;\rho) \d v.
\end{equation}
In the spatially homogeneous case we have $\rho=\rho(t=0)$. Since in traffic flow there is only a single collision invariant, the Maxwellian is parameterized by $\rho$. Mass conservation if fulfilled provided
$$
\mathcal{P}(v_*\to v | v^*;\rho) \geq 0, \quad \int_{\mathcal{V}} \mathcal{P}(v_*\to v | v^*;\rho) \d v = 1.
$$
The Maxwellian plays a role also in particular spatially non--homogeneous models. \revision{This is the case of the Bhatnagar--Gross--Krook (BGK) approximation~\cite{BGK1954} of the operator $Q[f,f]$ that we will consider from Section~\ref{sec:review:bgk} on.}

For reviews of kinetic models for traffic we refer to~\cite{albi2019vehicular,BellomoReview,piccoli2009ENCYCLOPEDIA}.

\subsection{A spatially homogeneous Boltzmann--type model for vehicular traffic} \label{sec:review:homogeneous}

In the case of vehicular traffic, we assume that the space $\mathcal{V}$ of the microscopic velocities of the vehicles is bounded $\mathcal{V}=[0,\vmax]$. We assume that $\mathcal{P}$ models the adaptation of the vehicles' speeds by binary car--to--car interaction where drivers react to the actions of the vehicle in front only. Interaction rules introduced in~\cite{PSTV2} are defined by:
\begin{equation} \label{eq:HtoHA}
v = 
\begin{cases}
\min\{v+\Delta v,\vmax\}, & \text{with probability } \ P(\rho) \\
v_*, & \text{with probability } \ 1-P(\rho) \ \text{ if } \ v_*\leq v^* \\
v^*, & \text{with probability } \ 1-P(\rho) \ \text{ if } \ v_*>v^*,
\end{cases}
\end{equation}
where $P:[0,\rmax] \to [0,1]$ is assumed to be a decreasing function of the density weighting acceleration and braking. The parameter $\Delta v>0$ will be referred to as acceleration parameter, and it allows to model the instantaneous but bounded acceleration. \revision{Binary interactions~\eqref{eq:HtoHA} are stochastic since vehicles' behavior is modeled by a probability of the local density $\rho$. A similar modeling approach in traffic is used by various authors, e.g.~see~\cite{coscia2007IJNM,DelitalaTosin2007,FermoTosin13}.}
The operator $\mathcal{P}$ summarizing the interaction rules~\eqref{eq:HtoHA} writes as
\begin{equation} \label{eq:rule}
\mathcal{P}(v_* \to v | v^*;\rho) =
\begin{cases} 
P(\rho) \, \delta_{\min\{ v_* + \Delta v, \vmax \}}(v) + (1-P(\rho)) \, \delta_{v_*}(v), & v_* \leq v^* \\ 
P(\rho) \, \delta_{\min\{ v_*+ \Delta v, \vmax \}} (v) + (1-P(\rho)) \, \delta_{v^*} (v), & v_* > v^* .
\end{cases}
\end{equation}
We recall that the continuity of the operator $\mathcal{P}$ across the line $v_*=v^*$ ensures well--posedness, see~\cite{PgSmTaVg3}.

For this particular model, the Maxwellian $\Maxw_f$ is a finite weighted sum of Dirac's distributions centered on speed values spaced by $\Delta v$ and is parameterized by the local density $\rho$. This structure of the stationary states is typically observed also in opinion dynamics models~\cite{BorraLorenzi2013,CanutoFagnaniTilli2008,CanutoFagnaniTilli12} based on a bounded confidence parameter.

\begin{mytheo}[Theorem 3.1 and Theorem 3.4 in~\cite{PSTV2}] \label{th:continuousEq}
	Let $P=P(\rho)$ be a given function of the density $\rho\in[0,\rmax]$ such that $P\in[0,1]$. Let $\{v_j\}_{j=1}^N$ be a set of equally spaced velocities in $\mathcal{V}=[0,\vmax]$, with $v_1=0$ and $v_N=\vmax$. Let $\Delta v=\frac{\vmax}{T}$, $T\in\mathbb{Z}^+$. Then, the probability distribution function
	\[
	\Maxw_f(v;\rho) = \sum_{j=1}^N f^{\infty}_j(\rho) \delta_{v_j}(v), \quad f^{\infty}_j > 0 \quad \forall\; j=1,\dots,N,
	\]
	with $\sum_{j=1}^N f_j^\infty(\rho)=\rho$, is the unique stable weak stationary solution of the model~\eqref{eq:homogBoltzmann}--\eqref{eq:rule} provided  $v_j=v_1+j\Delta v$, $j=1,\dots,N$, and
	\begin{equation*}
	\begin{aligned}
	f_1^\infty &= \begin{cases}
	0 & P \geq \tfrac12 \\
	\rho\frac{1-2P}{1-P} &\mbox{else}
	\end{cases} \\[1ex]
	f_j^\infty &= \begin{cases}
	0 & P \geq \tfrac12 \\ \frac{-2(1-P)\sum_{k=1}^{j-1} f_k^\infty+(1-2P)\rho+\sqrt{\left[(1-2P)\rho-2(1-P)\sum_{k=1}^{j-1}f_k^\infty\right]^2+4P(1-P)\rho f_{j-1}^\infty}}{2(1-P)} &\mbox{else}
	\end{cases}
	\end{aligned}
	\end{equation*}
	for $j=2,\dots,N-1$ and $f_N^\infty = \rho - \sum_{j=1}^{N-1} f_j^\infty$.
\end{mytheo}

\subsection{Macroscopic behavior via a Chapman--Enskog expansion} \label{sec:review:bgk}

The macroscopic properties of kinetic models can be studied using fundamental diagrams, i.e.~speed-- and flux--density relations at equilibrium defined in~\eqref{eq:macroQuantitiesEq}. In fact, comparing simulated and experimental fundamental diagrams allows to validate macroscopic features of typical traffic models. Instead, off--equilibrium \revision{phenomena} can be clarified using \revision{kinetic to macroscopic limits of perturbations around the equilibrium state, e.g.~via Chapman--Enskog expansion~\cite{Sopasakis,NelsonSopasakis} or Grad's moment method~\cite{Grad,Struchtrup2005}. In these regimes, a good approximation of the Boltzmann equation is provided by the BGK~\cite{BGK1954} model:}
\begin{equation} \label{eq:collisionKernel:bgk}
\begin{aligned}
	\partial_t f(t,x,v) + v \partial_x f(t,x,v) = \gamma \left( \Maxw_f(v;\rho) - f(t,x,v) \right).
\end{aligned}
\end{equation}
\revision{In the following, we focus on this class of kinetic equations. In particular the analysis is performed by using first order Chapman--Enskog expansions, namely we consider the regime where the kinetic distribution $f$ is a first order perturbation of the Maxwellian $\Maxw_f$, that is}
$$
	f(t,x,v) = \Maxw_f(v;\rho) + \frac{1}{\gamma} f^{(1)}(t,x,v), \quad \int_{\mathcal{V}} f^{(1)}(t,x,v) \mathrm{d}v = 0.
$$

\revision{It is possible to show that in this regime the evolution of the density $\rho$, cf.~\eqref{eq:density}, obtained from~\eqref{eq:collisionKernel:bgk}, is approximated by the advection--diffusion equation}
\begin{subequations} \label{eq:advectiondiffusion}
	\begin{align}
	\partial_t \rho(t,x) + \partial_x F(\rho) &= \frac{1}{\gamma} \partial_x \left( \mu(\rho) \partial_x \rho(t,x) \right), \label{eq:advdiffEq}\\
	\mu(\rho) &= \int_{\mathcal{V}} v^2 \partial_\rho \Maxw_f(v;\rho) \mathrm{d}v - F^\prime(\rho)^2. \label{eq:diffBGK}
	\end{align}
\end{subequations}
If the diffusion coefficient $\mu(\rho)$ is negative then the advection--diffusion equation is ill--posed and therefore has solutions with unbounded growth even starting from small perturbations of equilibrium states~\cite{HPRV20,SeiboldFlynnKasimovRosales2013,Seibold_BookICIAM}. \revision{This analysis is closely related to the investigation of the sub--characteristic condition in relaxation systems~\cite{Jin95therelaxation,Chen92hyperbolicconservation}. Observe that the convergence of~\eqref{eq:collisionKernel:bgk} to~\eqref{eq:advectiondiffusion}, as given in the previous considerations, occurs in the limit $\gamma\to\infty$. The parabolic term is thus a small perturbation term which provides information on small perturbations of equilibrium flow.} The following result \revision{summarizes} sufficient conditions on the Maxwellian leading to this macroscopic effect. Let us define the variance of the microscopic speeds at equilibrium as
\begin{equation} \label{eq:variance}
	\mathrm{Var}(\rho) = \int_{\mathcal{V}} (v-U(\rho))^2 \Maxw_f(v;\rho) \d v,
\end{equation}
where $U(\rho)$ is defined in~\eqref{eq:macroQuantitiesEq}

\begin{mytheo}[Proposition 2 in~\cite{HPRV20}] \label{th:diffusion}
	Assume that $\exists \, \widetilde{\rho} \in (0,\rmax)$ such that
	\begin{equation} \label{eq:conditions}
	\frac{\mathrm{d}}{\mathrm{d}\rho}F(\rho) < 0 \ \text{ and } \ \frac{\mathrm{d}}{\mathrm{d}\rho} \mathrm{Var}(\rho) < 0, \quad \forall\rho\in(\widetilde{\rho},\rmax).
	\end{equation}
	Then the diffusion coefficient~\eqref{eq:diffBGK} is negative in $\rho\in(\widetilde{\rho},\rmax)$ and thus~\eqref{eq:advdiffEq} is a forward--backward diffusion equation on $[0,\rmax]$.
\end{mytheo}

Conditions~\eqref{eq:conditions} are sufficient to \revision{ensure that}
$$
	\int_{\mathcal{V}} v^2 \partial_\rho \Maxw_f(v;\rho) \mathrm{d}v = \frac{\mathrm{d}}{\mathrm{d}\rho} \mathrm{Var}(\rho) + 2 U(\rho) \frac{\mathrm{d}}{\mathrm{d}\rho}F(\rho) - U^2(\rho) < 0,
$$
\revision{resulting in $\mu(\rho)<0$.} We observe that conditions~\eqref{eq:conditions} do occur in traffic flow. The first one is typically observed in the congested regime of experimental data, where the flow decreases as the density increases~\cite{Kerner1998}. The second condition is also realistic since we expect that the freedom in choosing a microscopic velocity reduces when traffic
becomes dense.
Both conditions are true for the kinetic model with the Maxwellian given in Theorem~\ref{th:continuousEq}, cf.~Figure~\ref{fig:measuresCmodel} in Section~\ref{sec:review:measures}. Therefore, where $\mu(\rho)<0$, the kinetic model produces unbounded growth of small perturbations of equilibrium states which propagates backwards since $F'(\rho)<0$. We say then that the BGK model \revision{is \emph{unstable}, i.e.~the flow is perturbed by sudden fluctuations~\cite{Sopasakis}. The sign of $\mu(\rho)$ provides information on the unstable regime according to the following definition.}

\begin{mydef} \label{def:stabilityBGK}
	\revision{We say that} a kinetic equation is stable in the first order Chapman--Enskog expansion if $\mu(\rho)\geq 0$, $\forall\,\rho\in[0,\rmax]$, weakly--unstable if $\mu(\rho)<0$ on an interval $(\alpha,\beta)$ properly contained in $[0,\rmax]$ and unstable if $\mu(\rho)<0$ on an interval $(\alpha,\beta)$ in which either $\alpha=0$ or $\beta=\rmax$. The interval $(\alpha,\beta)$ is said interval of instability and we denote by $\Gamma=\beta-\alpha$ its amplitude.
\end{mydef}

\revision{The inconsistency of the BGK model when applied to traffic flow problems is evident also by its macroscopic limit. In~\cite{HPRV20} we have proven that the macroscopic limit of~\eqref{eq:collisionKernel:bgk} is the Payne and Whitham model~\cite{Payne1971,Whitham1974} which is shown having several drawbacks~\cite{Daganzo1995}, such as an nonphysical speed propagation.}

\revision{For these reasons,} here we do not consider the BGK model~\eqref{eq:collisionKernel:bgk}. We focus\revision{, instead,} on the modified BGK--type equation introduced in~\cite{HPRV20} which was proven \revision{to converge to the Aw, Rascle and Zhang model in the macroscopic limit via Grad's moment method}:
\begin{equation} \label{eq:kineticW}
\partial_t g(t,x,w) + \partial_x \big[ (w-h(\rho)) g(t,x,w) \big] = \gamma \left( \Maxw_g(w;\rho) - g(t,x,w) \right).
\end{equation}
The model relies on a different interpretation of the microscopic velocities. In fact, here $g(t,x,w) : \mathbb{R}^+_0 \times \mathbb{R} \times \mathcal{W} \to \mathbb{R}^+_0$ is the distribution function of the \emph{desired} velocity $w\in\mathcal{W}:=[\wmin,+\infty)$, $\wmin>0$ is the minimum desired speed in free flow conditions. Compared to classical kinetic theory, the introduction of $g(t,x,w)$ allows us to consider another interpretation of traffic flow: a vehicle travels with the actual velocity $v$ because it cannot keep its own desired velocity $w$ due to traffic conditions. The deviation of $v$ from $w$ is measured by an increasing function of the local density. This is the pressure or hesitation function $h(\rho)$, such that $h'(\rho)>0$. In~\cite{HPRV20}, equation~\eqref{eq:kineticW} is derived as kinetic limit of the Bando and Follow--the--Leader model~\cite{Bando1995,FTL1961} and, therefore, $h(\rho)$ is the time derivative of the Follow--the--Leader term, see also~\cite{aw2002SIAP}. \revision{We point out that a kinetic model for the desired speeds was already considered by Paveri-Fontana in~\cite{paveri1975TR}, but desired and actual speeds were assumed to be independent. Instead, in~\cite{HPRV20} they are linked by the hesitation function.}

Despite the different viewpoint of the microscopic state, the macroscopic density at time $t$ and position $x$ is still defined by the zero--th moment of $g(t,x,w)$:
\begin{equation*} \label{eq:densityg}
\rho(t,x) = \int_{\mathcal{W}} g(t,x,w) \d w.
\end{equation*}
The distribution $\Maxw_g$ is the equilibrium distribution of the desired microscopic speeds and has to fulfill the requirement
$$
\int_{\mathcal{W}} \Maxw_g(w;\rho) \d w = \rho(t,x),
$$
and additionally
\begin{equation} \label{eq:MgRelation}
\frac{1}{\rho(t,x)} \int_{\mathcal{W}} w \Maxw_g(w;\rho) \d w = U(\rho) + h(\rho).
\end{equation}
\revision{It is important to notice that, thanks to~\eqref{eq:MgRelation}, the knowledge of $\Maxw_g$ is based on the knowledge of the classical Maxwellian $\Maxw_f$, i.e.~the one related to the actual microscopic velocity, by means of $U(\rho)$, cf.~\eqref{eq:macroQuantitiesEq}. We recall that $\Maxw_f$ comes out from the modeling of microscopic interactions of the spatially homogeneous kinetic model in Section~\ref{sec:review:homogeneous}.}

\revision{In~\cite{HPRV20}, the modified BGK model was also proven to be stable or weakly--stable in the congested phase of traffic, according to Definition~\ref{def:stabilityBGK}.}
In fact, performing a first order Chapman--Enskog expansion of the kinetic distribution $g(t,x,w)$,
it is possible to show that the BGK equation~\eqref{eq:kineticW} solves the advection--diffusion equation~\eqref{eq:advdiffEq} with
\begin{equation} \label{eq:newDiffusion}
\mu(\rho) = \int_{\mathcal{V}} v^2 \partial_\rho \Maxw_f(v;\rho) \d v - F^\prime(\rho)^2 - \rho h^\prime(\rho) F^\prime(\rho) + h^\prime(\rho) F(\rho).
\end{equation}
Observe that, compared to~\eqref{eq:diffBGK}, the diffusion coefficient~\eqref{eq:newDiffusion} contains an additional term which depends on the function $h(\rho)$. In particular, this term is non negative since $h$ and $U$ are an increasing and a non--increasing function of the density, respectively. Therefore, it is possible, for a given distribution $\Maxw_f$, to find a suitable $h(\rho)$ such that the model is stable or weakly--unstable according to Definition~\ref{def:stabilityBGK}.

For a detailed derivation of~\eqref{eq:kineticW} with related discussion we refer to~\cite{HPRV20}.

\section{Traffic stabilization through autonomous vehicles: a kinetic approach} \label{sec:autonomous}

One of the major goals of transportation engineering is the design of measures aimed to reduce unstable phenomena in traffic~\cite{WHO,SternEtAl2018}. Using a kinetic approach, we show that \revision{the presence} of autonomous vehicles reduces the speed variance among vehicles and the regime of densities where the model is weakly--unstable.

Similar studies, aimed to develop controlled dynamics and investigate the impact on speed variance reduction, have already been performed in the mathematical literature, e.g.~see~\cite{Piccoli2020}. The present work differs from previous ones since we do not use uncertainty quantification theory to distinguish between different classes of vehicles \revision{and we do not externally control autonomous cars. In fact, the autonomous vehicles modeled in this work are capable to provide enough information to prevent or diminish a stochastic response of drivers.}  

\subsection{Indicators of instability in traffic flow} \label{sec:review:measures}

We measure the instability of the flow of vehicles by taking into account the following two ``indicators''.
\begin{description}
	\item[Sign of the diffusion coefficient in the Chapman--Enskog expansion:] Stop and go waves are experimentally observed~\cite{Helbing,SternEtAl2018,DelleMonache2019} unstable phenomena occurring in some regimes of traffic as the result of the vehicles' inability to reach an equilibrium state. The key observation in~\cite{HPRV20,SeiboldFlynnKasimovRosales2013,Seibold_BookICIAM} is that relaxation models possess a phase transition determined by properties of the diffusion function arising in the relaxation or diffusive limit. For certain densities, where the diffusion function is positive, the solutions decay to equilibrium; while for larger densities, where the diffusion is negative, the solutions develop backward propagating traffic waves that can be regarded as models for stop and go waves.
	\item[Variance of microscopic speeds:] The high variability of vehicle speeds on a road is known to be responsible for the appearance of flow instabilities, e.g.~\cite{VadebyForsman2014,WHO}, and determines also the desire for lane--changing which is one of the major sources of risk, e.g.~\cite{FarooqJuhasz2020,HertyViscontiIFAC}. \revision{Close to equilibrium, the dominant term in the distribution function is the Maxwellian distribution, which is computed from the space homogeneous model.
	For this reason, here the variance of microscopic speeds
	is 
	analyzed at equilibrium 
	on the steady state of a spatially homogeneous kinetic model.}
\end{description}

\begin{figure}[!t]
	\centering
	\includegraphics[width=\textwidth]{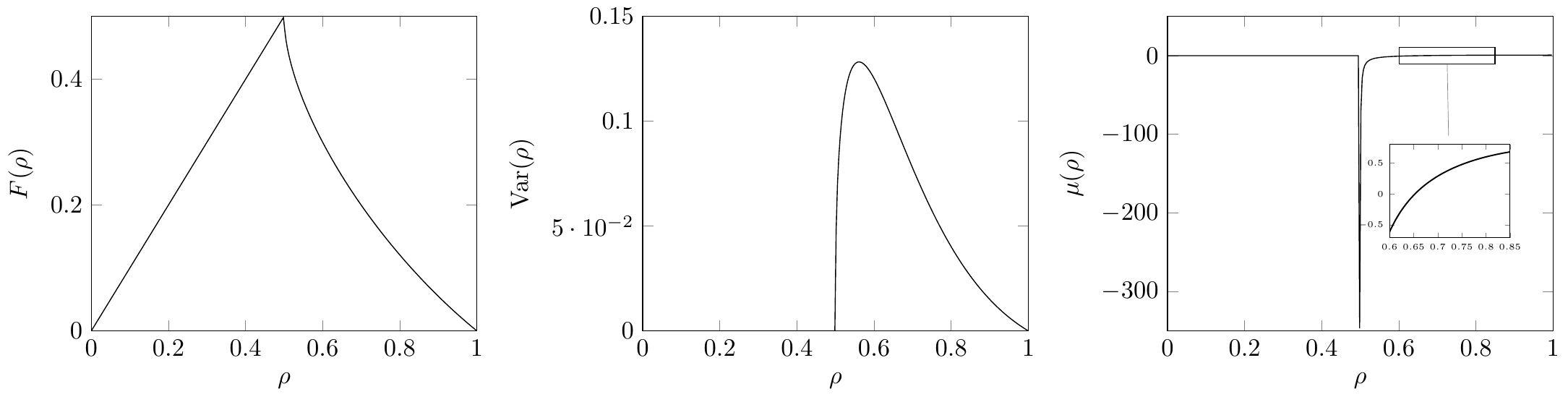}
	\caption{Left: flux--density diagram~\eqref{eq:macroQuantitiesEq}. Middle: variance of microscopic speeds at equilibrium~\eqref{eq:variance} as function of the density $\rho$. Right: diffusion coefficient~\eqref{eq:newDiffusion} in the diffusive limit of the BGK model~\eqref{eq:kineticW} as function of the density $\rho$. The plots are obtained using the Maxwellian of the spatially homogeneous model~\eqref{eq:homogBoltzmann}--\eqref{eq:rule} with $\Delta v=\frac{\vmax}{3}$. Here, $\rmax=1$ and $\vmax=1$.\label{fig:measuresCmodel}}
\end{figure}

We analyze the previous indicators on the kinetic model introduced in Section~\ref{sec:review}. In fact, the analysis of the two indicators is performed with the Maxwellian distribution of the spatially homogeneous model~\eqref{eq:homogBoltzmann}--\eqref{eq:rule}, described in Theorem~\ref{th:continuousEq}. We recall also that we consider the diffusion coefficient~\eqref{eq:newDiffusion}, obtained from the diffusive limit of the BGK model~\eqref{eq:kineticW}.

In Figure~\ref{fig:measuresCmodel} we show the flux--density diagram $\rho\mapsto F(\rho)$ at equilibrium (left panel) and the behavior of the two indicators of instability. In the middle panel, the variance of the microscopic speeds~\eqref{eq:variance} is computed at equilibrium. Instead, in the right panel we show the diffusion coefficient $\mu(\rho)$. We observe that the variance is zero in the free phase of traffic, here the regime corresponding to density values $\rho\leq 0.5$. In fact, as already noticed in Section~\ref{sec:review:homogeneous}, in this phase $\Maxw_f(v;\rho) = \delta_{\vmax}(v)$ and thus all vehicles travel with maximum speed and no speed variability is observed. As the density increases we notice a sudden increase of the speed variance, which reaches its maximum immediately after the critical density, and then decreases showing less speed variability when traffic becomes denser. Accordingly, we observe that the diffusion coefficient is zero for $\rho\leq 0$ which means that traffic is at equilibrium. Immediately after the critical density, the diffusion becomes large and negative and~\eqref{eq:advdiffEq} results in a backward advection--diffusion equation. In this regime the appearance of unstable waves is observed, with the property of being backward propagating~\cite{HPV_Book} since $F'(\rho)<0$. In addition, the numerical evidence \revision{shown in~\cite{HPRV20}} suggests that these waves are bounded thanks to the fact that the diffusion coefficient is negative in a proper domain $[\alpha,\beta]$, with $\alpha>0$ and $\beta<\rmax$. Thus, as pointed out in~\cite{HPRV20,SeiboldFlynnKasimovRosales2013}, instabilities produced in regimes where the model is weakly--unstable can be interpreted as stop and go waves.

\subsection{Microscopic interaction rules and the kinetic model with autonomous vehicles}

We study the impact of autonomous vehicles using a kinetic model that considers traffic as a mixture of two classes of vehicles, human and autonomous cars. Compared to~\cite{PgSmTaVg3}, here the two classes of vehicles have the same microscopic characteristics and differ in the microscopic interactions. The presence of two types of vehicles requires the definition of both self-- and cross--interactions.
We assume that human--driven vehicles are still characterized by the same rules presented in~\eqref{eq:HtoHA}. This assumption relies on the idea that regular vehicles are not able to distinguish between the two classes.

We introduce the labels $\AV$ and $\HV$ to identify the autonomous and human--driven vehicles, respectively. The density of the two classes of vehicles are $\rho_{\AV}=p\rho$ and $\rho_{\HV}=(1-p)\rho$, where $p\in[0,1]$ models the fixed penetration rate of autonomous vehicles, where $\rho\in[0,\rmax]$ is the total density of all cars.
As in Section~\ref{sec:review:homogeneous} we consider binary interactions between a vehicle, traveling with speed $v_*$, and a leading vehicle, traveling with speed $v^*$. As a result of the interaction, the vehicle with speed $v_*$ changes its speed in $v$ while the leading vehicle's speed remains unchanged. Thus, pre--collision states are $(v_*,v^*)$ and post--collision states are $(v,v^*)$.

\revision{We assume that autonomous vehicles are equipped with sensors that provide an instantaneous and precise description of the local conditions of the flow to their drivers, thus reducing their stochastic behavior. As a result of this information their drivers (human or computer) are aware of the presence of other autonomous cars ahead. Here, for the sake of illustrations, we take this feature to the extreme, i.e.~we assume that autonomous cars have no stochasticity in their decisions.
Since autonomous vehicles are able to distinguish between the two classes of vehicles, they can adapt their behavior based on the class of the interacting vehicle.} Then, in the case of self--interactions $\AV$--to--$\AV$ we assign
\begin{equation} \label{eq:AtoA}
	v = \min\{v+\Delta v,\hat{u}(\rho)\},
\end{equation}
while in the case of cross--interactions $\AV$--to--$\HV$ we impose
\begin{equation} \label{eq:AtoC}
v = 
\begin{cases}
\min\{v+\Delta v,\hat{u}(\rho)\}, & \text{if } \ \rho<\bar{\rho} \\
v_*, & \text{if } \ \rho\geq \bar{\rho} \ \text{ and } \ v_*\leq v^* \\
v^*, & \text{if } \ \rho\geq \bar{\rho} \ \text{ and } v_*>v^*.
\end{cases}
\end{equation}
In the previous rules, $\hat{u}(\rho)$ is a velocity and is a given function of the density $\rho$ with properties
\begin{align*}
\hat{u}(0)=\vmax, \quad \hat{u}(\rmax)=0, \quad \hat{u}^\prime(\rho)<0,
\end{align*}
whereas $\bar{\rho}\in[0,\rmax]$. Note that, unlike~\eqref{eq:HtoHA}, we assume that autonomous cars have a deterministic behavior. \revision{The speed $\hat{u}(\rho)$ is a desired speed: it could be chosen as the speed that permits a vehicle can arrest in the distance which avoids the collision with the vehicle ahead. For another possible choice see Section~\ref{sec:analysis}.}

As in~\cite{PgSmTaVg3} we are considering a multi-class model described by the coupled Boltzmann--type equations
\begin{equation} \label{eq:twoDistrModel}
\begin{aligned}
\partial_t \fh(t,v) & = \gamma Q[\fh,\fh](t,v) + \gamma Q[\fh,\fa](t,v)  \\
\partial_t \fa(t,v) & = \gamma Q[\fa,\fa](t,v) + \gamma Q[\fa,\fh](t,v)
\end{aligned}
\end{equation}
describing the evolution of the mass distribution functions $\fh,\fa:\mathbb{R}^+_0\times\mathcal{V} \to \mathbb{R}^+_0$ of the human--driven and autonomous vehicles, respectively, which satisfy
$$
	\int_{\mathcal{V}} \fh(t,v) \d v = \rho_{\HV}, \quad \int_{\mathcal{V}} \fa(t,v) \d v = \rho_{\AV}.
$$
Thus, generalizing the single--class collision kernel~\eqref{eq:homogBoltzmann}, for each $\mathsf{p},\mathsf{q}\in\{\HV,\AV\}$ we have
\begin{equation} \label{eq:twoDistrKernel}
	 Q[f_\mathsf{p},f_\mathsf{q}](t,v) = \int_\mathcal{V} \int_\mathcal{V} \mathcal{P}_{\mathsf{p}\mathsf{q}}(v_*\to v | v^*;\rho) f_\mathsf{p}(t,v_*) f_\mathsf{q}(t,v^*) \d v_* \d v^* - \rho_\mathsf{q} f_\mathsf{p}(t,v).
\end{equation}
For $\mathsf{p}=\mathsf{q}=\HV$ or $\mathsf{p}=\HV$ and $\mathsf{q}=\AV$, the probability operator $\mathcal{P}_{\mathsf{p}\mathsf{q}}$ is given by~\eqref{eq:rule}. Instead, considering the interaction rules~\eqref{eq:AtoA}--\eqref{eq:AtoC} for the autonomous vehicles, we have
\begin{equation} \label{eq:operatorAA}
	\mathcal{P}_{\mathsf{p}\mathsf{q}}(v_*\to v | v^*;\rho) = \delta_{\min\{ v_* + \Delta v, \hat{u}(\rho) \}}(v)
\end{equation}
if $\mathsf{p}=\mathsf{q}=\AV$, and
\begin{equation} \label{eq:operatorAC}
	\mathcal{P}_{\mathsf{p}\mathsf{q}}(v_*\to v | v^*;\rho) =
	\begin{cases}
		H(\bar{\rho}-\rho) \delta_{\min\{ v_* + \Delta v, \hat{u}(\rho) \}}(v) + H(\rho-\bar{\rho}) \delta_{v_*}(v), & v_* \leq v^* \\
		H(\bar{\rho}-\rho) \delta_{\min\{ v_* + \Delta v, \hat{u}(\rho) \}}(v) + H(\rho-\bar{\rho}) \delta_{v^*}(v), & v_* > v^*
	\end{cases}
\end{equation}
if $\mathsf{p}=\AV$ and $\mathsf{q}=\HV$, where $H(\cdot)$ is the Heaviside function.

Inserting~\eqref{eq:rule}--\eqref{eq:operatorAA}-\eqref{eq:operatorAC} into~\eqref{eq:twoDistrKernel} we obtain the following explicit expression of the collision kernels. For $\mathsf{p}=\HV, \ \mathsf{q}\in\{\HV,\AV\}$
\begin{align*}
	Q[f_\mathsf{p},f_\mathsf{q}](t,v) =& \rho_\mathsf{q} P(\rho) \left[ \delta_{\vmax}(v) \int_{\vmax-\dv}^{\vmax} f_\mathsf{p}(t,v_*) \d v_* + \int_0^{\vmax-\dv} \delta_{v-\dv}(v_*) f_\mathsf{p}(t,v_*) \d v_* \right]\\
	& + (1-P(\rho)) \left[ f_\mathsf{p}(t,v)  \int_{v}^{\vmax} f_\mathsf{q}(t,v^*) \d v^* + f_\mathsf{q}(t,v)  \int_{v}^{\vmax} f_\mathsf{p}(t,v_*) \d v_* \right] - \rho_{\mathsf{q}} f_\mathsf{p}(t,v) \\
	=& \rho_\mathsf{q} P(\rho) \left[ \delta_{\vmax}(v) \int_{\vmax-\dv}^{\vmax} f_\mathsf{p}(t,v_*) \d v_* + H(v-\Delta v) f_\mathsf{p}(t,v-\Delta v) \right]\\
	& + (1-P(\rho)) \left[ f_\mathsf{p}(t,v)  \int_{v}^{\vmax} f_\mathsf{q}(t,v^*) \d v^* + f_\mathsf{q}(t,v)  \int_{v}^{\vmax} f_\mathsf{p}(t,v_*) \d v_* \right] - \rho_{\mathsf{q}} f_\mathsf{p}(t,v),
\end{align*}
whereas
\begin{align*}
	Q[\fa,\fa](t,v) = & \rho_\AV \left[\delta_{\hat{u}(\rho)}(v) \int_{\hat{u}(\rho)-\dv}^{\vmax} \fa(t,v_*) \d v_* + \int_{0}^{\hat{u}(\rho)-\dv} \delta_{v-\dv}(v_*) \fa(t,v_*) \d v_*\right] - \rho_\AV \fa(t,v) \\
	= & \rho_\AV \left[\delta_{\hat{u}(\rho)}(v) \int_{\hat{u}(\rho)-\dv}^{\vmax} \fa(t,v_*) \d v_* + \chi_{[\dv,\hat{u}(\rho)]}(v) \fa(t,v-\dv)\right] - \rho_\AV \fa(t,v),
\end{align*}
and, finally,
\begin{align*}
	Q[\fa,\fh](t,v) = & \rho_{\HV} H(\bar{\rho}-\rho) \left[ \delta_{\hat{u}(\rho)}(v) \int_{\hat{u}(\rho)-\dv}^{\vmax} \fa(t,v_*) \d v_* + \int_0^{\hat{u}(\rho)-\dv} \delta_{v-\dv}(v_*) \fa(t,v_*) \d v_* \right]\\
	& + H(\rho-\bar{\rho}) \left[ \fa(t,v)  \int_{v}^{\vmax} \fh(t,v^*) \d v^* + \fh(t,v)  \int_{v}^{\vmax} \fa(t,v_*) \d v_* \right] - \rho_\HV \fa(t,v) \\
	= & \rho_{\HV} H(\bar{\rho}-\rho) \left[ \delta_{\hat{u}(\rho)}(v) \int_{\hat{u}(\rho)-\dv}^{\vmax} \fa(t,v_*) \d v_* + \chi_{[\dv,\hat{u}(\rho)]}(v) \fa(t,v-\dv) \right]\\
	& + H(\rho-\bar{\rho}) \left[ \fa(t,v)  \int_{v}^{\vmax} \fh(t,v^*) \d v^* + \fh(t,v)  \int_{v}^{\vmax} \fa(t,v_*) \d v_* \right] - \rho_\HV \fa(t,v).
\end{align*}
The two--distribution model~\eqref{eq:twoDistrModel} satisfies mass conservation.

\subsection{The single--distribution model and its qualitative properties} \label{sec:analysis}

The complexity of~\eqref{eq:twoDistrModel} does not allow for analytical estimates. We therefore derive a kinetic equation for the evolution of the total distribution $f(t,x,v)$ of vehicles. This requires to assume a relation between $f$ and $\fh$, $\fa$, respectively. \revision{If the penetration rate is small, we assume that $f_{\HV}$ is the dominant term in the total distribution $f$. Thus,} we make the following Ansatz:
\begin{equation} \label{eq:closure1}
	\fh(t,v) = \frac{\rho_\HV}{\rho}f(t,v), \quad \fa(t,v) = \frac{\rho_\AV}{\rho}f(t,v),
\end{equation}
which implies
\begin{equation*}
	\int_{\mathcal{V}} f(t,v) \d v = \rho_{\HV} + \rho_{\AV} = \rho.
\end{equation*}
Then, we obtain
\begin{equation} \label{eq:singleDistrModel}
	\partial_t f(t,v) = \partial_t \left( \fa(t,v) + \fh(t,v) \right) =  \sum_{\mathsf{p},\mathsf{q}\in\{\HV,\AV\}} \gamma Q[f_\mathsf{p},f_\mathsf{q}](t,v) = \gamma Q[f,f](t,v),
\end{equation}
where the collision operator is
\begin{equation} \label{eq:singleDistrKernel}
\begin{aligned}
	Q[f,f](t,v) =& \rho_{\HV} P(\rho) \left[ \delta_{\vmax}(v) \int_{\vmax-\dv}^{\vmax}  f(t,v_*) \d v_* + H(v-\dv) f(t,v-\dv) \right] \\
	& + 2 \frac{\rho_{\HV}}{\rho} \left( 1-P(\rho) + \frac{\rho_{\AV}}{\rho} H(\rho-\bar{\rho}) \right) f(t,v) \int_v^{\vmax} f(t,w) \d w \\
	& + \frac{\rho_{\AV}}{\rho} \left( \rho_{\AV} + \rho_{\HV} H(\bar{\rho}-\rho) \right) \left[ \delta_{\hat{u}(\rho)}(v) \int_{\hat{u}(\rho)-\dv}^{\vmax} f(t,v_*) \d v_* + \chi_{[\dv,\hat{u}(\rho)]}(v) f(t,v-\dv) \right] \\
	& - \rho f(t,v).
\end{aligned}
\end{equation}


Under simplifying assumptions it is possible to derive insights on the qualitative behavior of macroscopic solutions. Assume from now on
\begin{equation} \label{eq:unboundedDV}
	\Delta v = \vmax.
\end{equation}
Note that this choice does not imply a $2$--velocity model at equilibrium because of the microscopic interactions of the autonomous vehicles which depend on the velocity $\hat{u}(\rho)\in[0,\vmax]$. The model~\eqref{eq:singleDistrModel}--\eqref{eq:singleDistrKernel} reduces to
\begin{equation} \label{eq:simplifiedModel}
\begin{aligned}
	\partial_t f(t,v) =& \gamma\rho\rho_{\HV} P(\rho) \delta_{\vmax}(v) + 2 \gamma \frac{\rho_{\HV}}{\rho} \left(1-P(\rho)+\frac{\rho_{\AV}}{\rho}H(\rho-\bar{\rho})\right) f(t,v) \int_v^{\vmax} f(t,w) \d w \\
	& + \gamma\rho_{\AV} \left(\rho_{\AV}+\rho_{\HV}H(\bar{\rho}-\rho)\right) \delta_{\hat{u}(\rho)}(v) - \gamma \rho f(t,v).
\end{aligned}
\end{equation}
For any $k\geq 0$, the $\rho$--normalized $k$--th moment is
\begin{equation*}
	m_k(t;\rho) = \frac{1}{\rho} \int_{\mathcal{V}} v^k f(t,v) \d v.
\end{equation*}
Then, $m_1(t;\rho) = u(t)$, i.e.~the first normalized moment corresponds to the mean speed of the flow, and the variance of the microscopic speeds is
\begin{equation*}
	\mathrm{Var}(t;\rho) = m_2(t;\rho) - m_1(t;\rho)^2.
\end{equation*}
The evolution of the $k$--th moment is given by the following ordinary differential equation:
\begin{equation} \label{eq:kMomEq}
\begin{aligned}
	\frac{\mathrm{d}}{\mathrm{d}t} m_k(t;\rho) = & \gamma\rho_{\HV}\vmax^k P(\rho) + 2\gamma\frac{\rho_{\HV}}{\rho^2} \left(1-P(\rho)+\frac{\rho_{\AV}}{\rho}H(\rho-\bar{\rho})\right) \int_{\mathcal{V}} v^k f(t,v) \d v \int_v^{\vmax} f(t,w) \d w \\
	&+ \gamma \frac{\rho_{\AV}}{\rho} \left(\rho_{\AV}+\rho_{\HV}H(\bar{\rho}-\rho)\right) \hat{u}^k(\rho) - \gamma \rho m_k(t;\rho).
\end{aligned}
\end{equation}

\paragraph{Behavior of the variance.} The optimal choice of the velocity $\hat{u}(\rho)$ can be \revision{defined as the speed which minimizes the variance of microscopic speeds. Now} the variance is given by
\begin{align*}
\frac{\mathrm{d}}{\mathrm{d}t} \mathrm{Var}(t;\rho) =& -\gamma \rho \mathrm{Var}(t;\rho) + \gamma \frac{\rho_{\AV}}{\rho} \left( \rho_{\AV}+\rho_{\HV} H(\bar{\rho}-\rho) \right) \left( \hat{u}(\rho)^2-2\hat{u}(\rho)m_1(t;\rho) \right) \\ 
&+\gamma\rho_{\HV} \vmax P(\rho) \left(\vmax - 2m_1(t;\rho) \right) +\gamma \rho m_1(t;\rho)^2 \\
&+ 2\gamma\frac{\rho_{\HV}}{\rho^2} \left( 1-P(\rho) + \frac{\rho_{\AV}}{\rho} H(\rho-\bar{\rho}) \right) \int_{\mathcal{V}} (v^2-2m_1 v) f(t,v) \mathrm{d}v \int_{v}^{\vmax} f(t,w) \mathrm{d}w.
\end{align*}
The second term is the only one depending on $\hat{u}(\rho)$. This suggests to choose $\hat{u}(\rho) = m_1(t;\rho)$ in order to minimize the variance at equilibrium.

\paragraph{Capacity drop in the equilibrium flux.} We discuss the role of $\bar{\rho}$ in the equilibrium dynamics. We compute the fundamental diagram $F(\rho) = \rho U(\rho)$ from~\eqref{eq:firstMomEq}, where $F$ and $U$ are defined in~\eqref{eq:macroQuantitiesEq}. Then, at equilibrium we have
\begin{equation*}
F(\rho) = 
\begin{cases}
\displaystyle{\rho_{\HV} \vmax P(\rho) + \rho_{\AV} \hat{u}(\rho) + 2 \frac{\rho_{\HV}}{\rho^2} (1-P(\rho)) X(\rho)}, & \rho<\bar{\rho} \\[2ex]
\displaystyle{\rho_{\HV} \vmax P(\rho) + \frac{\rho_{\AV}^2}{\rho} \hat{u}(\rho) + 2 \frac{\rho_{\HV}}{\rho^2} (1-P(\rho) + \frac{\rho_{\AV}}{\rho}) X(\rho)}, & \rho>\bar{\rho},
\end{cases}
\end{equation*}
where $X(\rho):=\int_{\mathcal{V}} v \Maxw_{f}(v;\rho) \d v \int_v^{\vmax} \Maxw_{f}(w;\rho) \d w$.
Hence, the fundamental diagram may have a discontinuity at $\rho=\bar{\rho}$.

If we choose $\hat{u}(\rho) = U(\rho)$ which, as we have seen, minimizes the variance of microscopic speeds at equilibrium, we have $\rho \hat{u}(\rho) = F(\rho)$. Using $\rho_{\HV}=\rho-\rho_{\AV}$, the equilibrium flux is
\begin{equation*}
F(\rho) = 
\begin{cases}
\displaystyle{\rho \vmax P(\rho) + 2 \frac{1-P(\rho)}{\rho} X(\rho)}, & \rho<\bar{\rho} \\[2ex]
\displaystyle{\frac{\rho^2}{\rho+\rho_{\AV}} \vmax P(\rho) + \frac{2}{\rho+\rho_{\AV}}\left(1-P(\rho) + \frac{\rho_{\AV}}{\rho}\right) X(\rho)}, & \rho>\bar{\rho}.
\end{cases}
\end{equation*}
We observe that the flux is not explicitly influenced by the penetration rate of the autonomous vehicles if $\rho<\bar{\rho}$. Conversely, for $\rho>\bar{\rho}$ the behavior of the flux depends on the density of autonomous vehicles on the road. In particular, the height of the jump at $\rho=\bar{\rho}$ is
\begin{align*}
F(\bar{\rho}^-)-F(\bar{\rho}^+) = \frac{(\rho-\rho_{\AV})|\rho^2\vmax-2X|P(\rho)}{\rho(\rho+\rho_{\AV})} .
\end{align*}
If $\rho_{\AV} \to \rho$ the jump of the equilibrium flux at $\rho=\bar{\rho}$ is zero, namely the fundamental diagram is continuous. The capacity drop of the flux at $\rho=\bar{\rho}$ decreases if the number of autonomous vehicles on the road increases. However the penetration rate of autonomous vehicles is typically low, and so $\rho_{\AV} \to \rho$ is difficult to observe. Thus, this analysis also suggests that the capacity drop in traffic can be minimized with the presence of autonomous vehicles as $\bar{\rho}\to1$.

\paragraph{Behavior of the mean speed.} The evolution of the mean speed is given by the following ordinary differential equation:
\begin{equation} \label{eq:firstMomEq}
\begin{aligned}
	\frac{\mathrm{d}}{\mathrm{d}t} m_1(t;\rho) = & \gamma\rho_{\HV}\vmax P(\rho) + 2\gamma\frac{\rho_{\HV}}{\rho^2} \left(1-P(\rho)+\frac{\rho_{\AV}}{\rho}H(\rho-\bar{\rho})\right) \int_{\mathcal{V}} v f(t,v) \d v \int_v^{\vmax} f(t,w) \d w \\
	&+ \gamma\frac{\rho_{\AV}}{\rho} \left(\rho_{\AV}+\rho_{\HV}H(\bar{\rho}-\rho)\right) \hat{u}(\rho) - \gamma \rho m_1(t;\rho).
\end{aligned}
\end{equation}
Note that
\begin{equation} \label{eq:x}
	X(\rho):=\int_{\mathcal{V}} v f(t,v) \d v \int_v^{\vmax} f(t,w) \d w \leq \rho^2 m_1(t;\rho).
\end{equation}
Then, for $\rho>\bar{\rho}$ we have
\begin{equation*}
\begin{aligned}
	\frac{\mathrm{d}}{\mathrm{d}t} m_1(t;\rho) \leq & \gamma \rho_{\HV}\vmax P(\rho) + \gamma\frac{\rho_{\AV}^2}{\rho} \hat{u}(\rho) + \gamma a(p,\rho) m_1(t;\rho) \\
	a(p,\rho) =& \rho(1-2p^2-2(1-p)P(\rho)).
\end{aligned}
\end{equation*}
Then, if $a(p,\rho)<0$ the first moment decreases and the asymptotic value is larger when autonomous vehicles are present. This happens if $P(\rho)>\frac{1-2p^2}{2(1-p)}$. Since $P$ is a decreasing function of the density and the penetration rate $p$ is typically low, we observe that $P(\rho)>\frac{1-2p^2}{2(1-p)}$ is verified for low density regimes. For instance, let us take $P(\rho)=1-\frac{\rho}{\rmax}$, then $a(p,\rho)<0$ if $\frac{\rho}{\rmax}<\frac{1-p}{2}$. In this regime, autonomous vehicles allow to have a larger speed of the flow.
A similar consideration holds if $\rho<\bar{\rho}$. In this case the evolution of the mean speed is given by
\begin{equation*}
\begin{aligned}
	\frac{\mathrm{d}}{\mathrm{d}t} m_1(t;\rho) \leq & \gamma \rho_{\HV} \vmax P(\rho) + \gamma \rho_{\AV} \hat{u}(\rho) + \gamma b(p,\rho) m_1(t;\rho) \\
	b(p,\rho) &= \rho(1-2p-2(1-p)P(\rho)),
\end{aligned}
\end{equation*}
and $b(p,\rho)<0$ is verified if $\frac{\rho}{\rmax}<\frac{1}{2(1-p)}$ when $P(\rho)=1-\frac{\rho}{\rmax}$.

\begin{figure}
	\centering
	\includegraphics[width=0.5\textwidth]{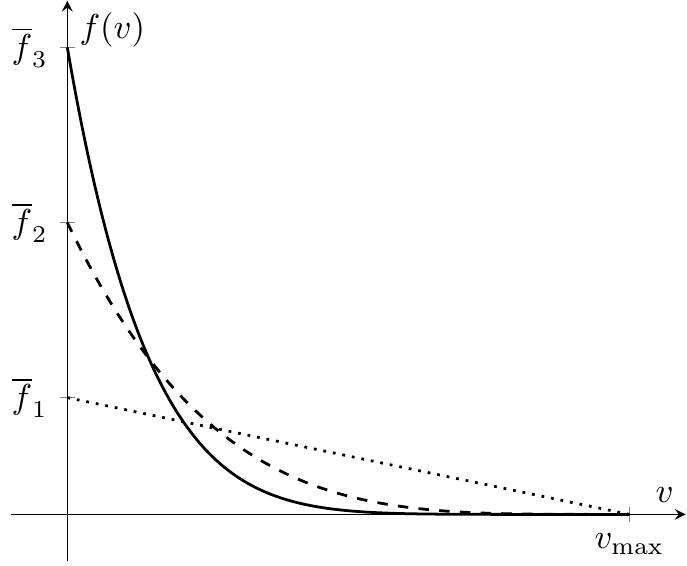}
	\caption{Schematic representation of the Ansatz~\eqref{eq:fShape} for the kinetic distribution with $\rho=0.8\rmax$ and $\delta_1=1$ (dotted line), $\delta_2=4$ (dashed line), $\delta_3=7$ (solid line). The values $\overline{f}_i$, $i=1,2,3$, are found in order to satisfy mass conservation.\label{fig:fShape}}
\end{figure}

The previous results are limited to a particular regime of traffic due to the rough approximation of $X$, cf.~\eqref{eq:x}. A more general result can be obtained with a finer approximation of $X$ which is based on an assumption on the shape of the distribution function $f$. Since $\bar{\rho}$ plays the role of a critical density, let us take
\begin{equation} \label{eq:fShape}
	f(v) \approx \begin{cases}\rho\frac{\delta+1}{\vmax^2} (\vmax-v)^\delta, & \rho>\bar{\rho} \\
	\frac{\delta+1}{\vmax^{\delta+1}} \rho v^\delta, & \rho<\bar{\rho}	\end{cases}
\end{equation}
where $\delta>1$, see Figure~\ref{fig:fShape}. This choice relies on the simple idea that in low density regimes the distribution function $f$ is concentrated on high velocities. Conversely for high density regimes. With~\eqref{eq:fShape}, for $\rho>\bar{\rho}$ we have
$$
	X \approx \frac{\rho^2 \vmax^{2\delta -1}}{2(2\delta+3)}.
$$
Then, using also that $P(\rho)\leq 1-P(\rho)$ if $\rho>\bar{\rho}$ and high, from~\eqref{eq:firstMomEq} we obtain
$$
	\begin{aligned}
	\frac{\mathrm{d}}{\mathrm{d}t} m_1(t;\rho) \leq & \gamma\rho_{\HV} \vmax (1-P(\rho)) \left( 1 + \frac{\vmax^{2\delta-2}}{2\delta+3} \right) + \gamma \frac{\rho_{\AV}}{\rho} \left( \rho_{\AV} \hat{u}(\rho) + \rho_{\HV} \frac{\vmax^{2\delta-1}}{2\delta+3} \right) - \gamma \rho m_1(t;\rho).
	\end{aligned}
$$
Instead, for $\rho<\bar{\rho}$ we have
$$
X \approx \left(\frac{\delta+1}{\delta+2}-\frac{\delta+1}{2\delta+3}\right)\rho^2\vmax.
$$
Then, using also that $P(\rho)\geq 1-P(\rho)$ if $\rho<\bar{\rho}$ and low, from~\eqref{eq:firstMomEq} we obtain
$$
\frac{\mathrm{d}}{\mathrm{d}t} m_1(t;\rho) \leq \gamma \rho_{\HV} \vmax P(\rho) \left(1 + 2\frac{\delta+1}{\delta+2}-\frac{\delta+1}{2\delta+3}\right) + \gamma\rho_{\AV} \hat{u}(\rho)-\gamma\rho m_1(t;\rho).
$$
We observe that the presence of autonomous vehicles allows to obtain a higher first moment at equilibrium and that this result does not depend on a particular density regime.

\paragraph{Diffusion coefficient in the Chapman--Enskog expansion.}

Assuming that there exists a steady state solution to~\eqref{eq:singleDistrModel}--\eqref{eq:singleDistrKernel}, i.e. a kinetic distribution $\Maxw_{f}(v;\rho)$ such that $Q[\Maxw_{f},\Maxw_{f}]=0$, we postulate that the spatially non--homogeneous dynamics of autonomous and human--driven vehicles are described by the modified BGK structure~\eqref{eq:kineticW}. The difference with respect to the model with human--driven vehicles only is in the Maxwellian distribution. It is difficult to determine an analytical formulation for $\Maxw_{f}(v;\rho)$. However, the same Chapman--Enskog expansion discussed in Section~\ref{sec:review:bgk} applies to the case of a mixture of vehicles. Considering small perturbations of the kinetic distribution $f$ around the Maxwellian, we lead to~\eqref{eq:advdiffEq} where $\mu(\rho)$ is again as in~\eqref{eq:newDiffusion}, but with a different Maxwellian. This result is used in the numerical simulations in order to show the behavior of the sign of the diffusion coefficient.

\subsection{Numerical simulations}

In this section we numerically investigate the behavior of the single--distribution model for autonomous and human--driven vehicles including a more general setting than~\eqref{eq:unboundedDV}. To this end, as in several kinetic models, we employ the Nanbu--like asymptotic method~\cite{Nanbu} which we reformulate in Algorithm~\ref{alg:Nanbu} in Appendix~\ref{app} for the spatially homogeneous single--distribution model~\eqref{eq:singleDistrModel}. For further details we refer also to~\cite{AlbiPareschi,HertyKlarPareschi,PareschiToscaniBOOK}. Algorithm~\ref{alg:Nanbu} is employed to compute the stationary states of the kinetic model. The parameters of the Nanbu algorithm are taken as $N=20000$ particles and $M=200$ iterations, and it is performed on $50$ equally spaced values of densities in the interval $[0.01,0.99]$. The maximum density and velocity are normalized, i.e.~$\rmax=\vmax=1$, and the acceleration parameter is $\Delta v=\frac{1}{3}$. The initial velocities uniformly distributed in $[0,1]$, namely we consider $f(t=0,v)=\chi_{[0,1]}(v)$ and, due to the analysis in Section~\ref{sec:analysis}, we will always take $\hat{u}(\rho)$ as the mean speed, i.e.~$\hat{u}(\rho)=\frac{1}{\rho}\int_{\mathcal{V}} v f(t,v) \d v$. 

\begin{figure}[t!]
	\centering
	\includegraphics[width=\textwidth]{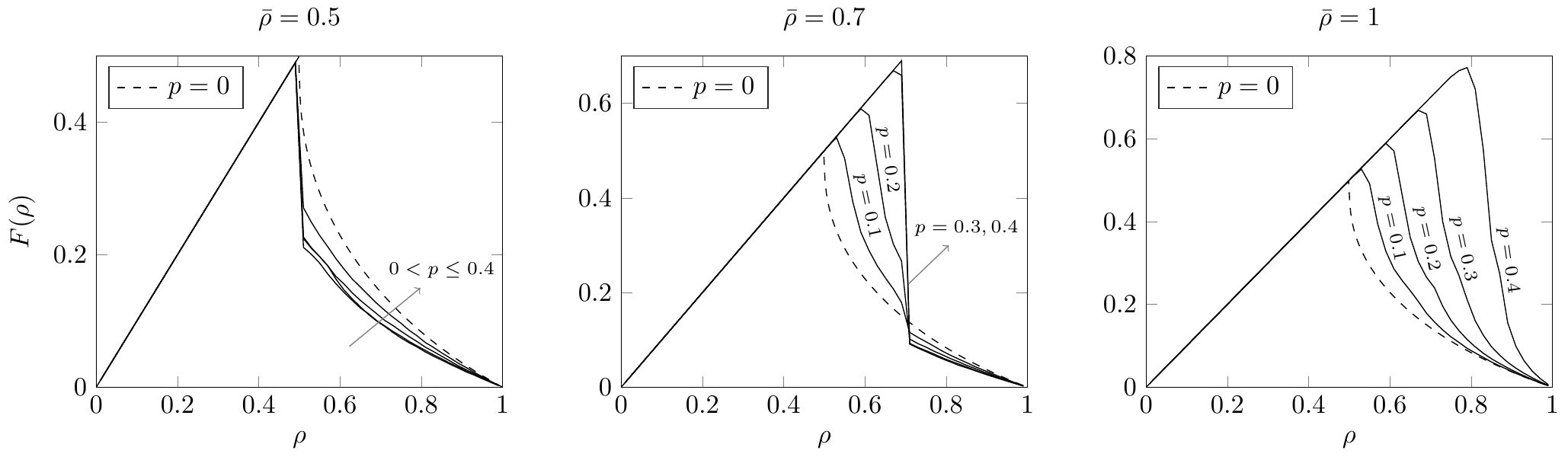}
	\caption{Fundamental diagrams of the single--distribution model~\eqref{eq:singleDistrModel} for autonomous and human--driven vehicles for three choices of the parameter $\bar{\rho}$. The different diagrams refer to different penetration rates, $p\leq 0.4$. Here $\vmax=1=\rmax=1$, $\Delta v=\frac13$ and $\hat{u}(\rho)=\frac{1}{\rho}\int_{\mathcal{V}} v f(t,v) \d v$.\label{fig:equilibriaCA}}
\end{figure}

In Figure~\ref{fig:equilibriaCA} we observe the fundamental diagrams, i.e.~the density--flux diagram at equilibrium, of the single--distribution model~\eqref{eq:singleDistrModel}. We consider five values for the penetration rate with $p\leq 0.4$, and three values of the parameter $\bar{\rho}$, namely $\bar{\rho}=0.5,0.7,1$. We recall that $\bar{\rho}$ determines the behavior of the autonomous vehicles when interacting with human--driven vehicles. In particular, $\bar{\rho}=1$ means that autonomous vehicles do not distinguish between autonomous and human--driven vehicles. Furthermore, the previous analysis, performed in the simplified setting $\Delta v=\vmax$, showed that $\bar{\rho}$ plays the role of critical density where a capacity drop in the fundamental diagram occurs. This behavior is observed also with a different choice of the acceleration parameter, namely with $\Delta v=\frac13$. Instead, no capacity drop of the density--flux diagram appears when $\bar{\rho}=1$ as expected. Also, the choice $\bar{\rho}=1$ provides the typical behavior of the flux we expect when we increase the percentage of autonomous vehicles on the road: the flux increases monotonically with respect to $p$, showing that a large penetration rate of autonomous vehicles allows for a higher flux and to move the congested regime, where we observe a decrease of the flux, to higher density values.

\begin{figure}[t!]
	\centering
	\includegraphics[width=\textwidth]{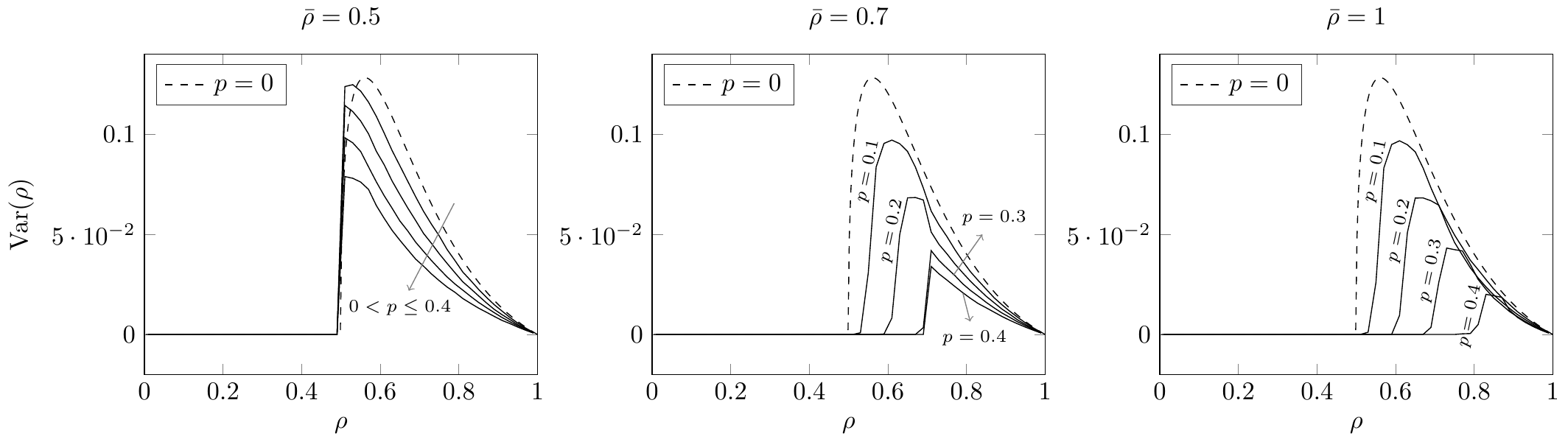}
	\caption{Variance of microscopic speeds at equilibrium of the single--distribution model~\eqref{eq:singleDistrModel} for autonomous and human--driven vehicles. Three choices of the parameter $\bar{\rho}$ and different penetration rates $p$ are considered. Here $\vmax=1=\rmax=1$, $\Delta v=\frac13$ and $\hat{u}(\rho)=\frac{1}{\rho}\int_{\mathcal{V}} v f(t,v) \d v$.\label{fig:varianceCA}}
\end{figure}

Now, we move the focus on the analysis of the instability measures defined in Section~\ref{sec:review:measures}. Firstly, we discuss the variance of microscopic speeds at equilibrium. In Figure~\ref{fig:varianceCA} we show the variance~\eqref{eq:variance}, as function of the density $\rho$, of the single--distribution model~\eqref{eq:singleDistrModel} of autonomous and human--driven vehicles. Also in this case, we compare the results for three values of the parameter $\bar{\rho}$ and with several values of the penetration rates. Observe that in all cases increasing the penetration rate causes a reduction of the speed variability at equilibrium. Furthermore, for $\bar{\rho}>0.5$, the maximum value of the variance moves towards larger density regimes which are typically not observed in real traffic.

\begin{figure}[t!]
	\centering
	\includegraphics[width=\textwidth]{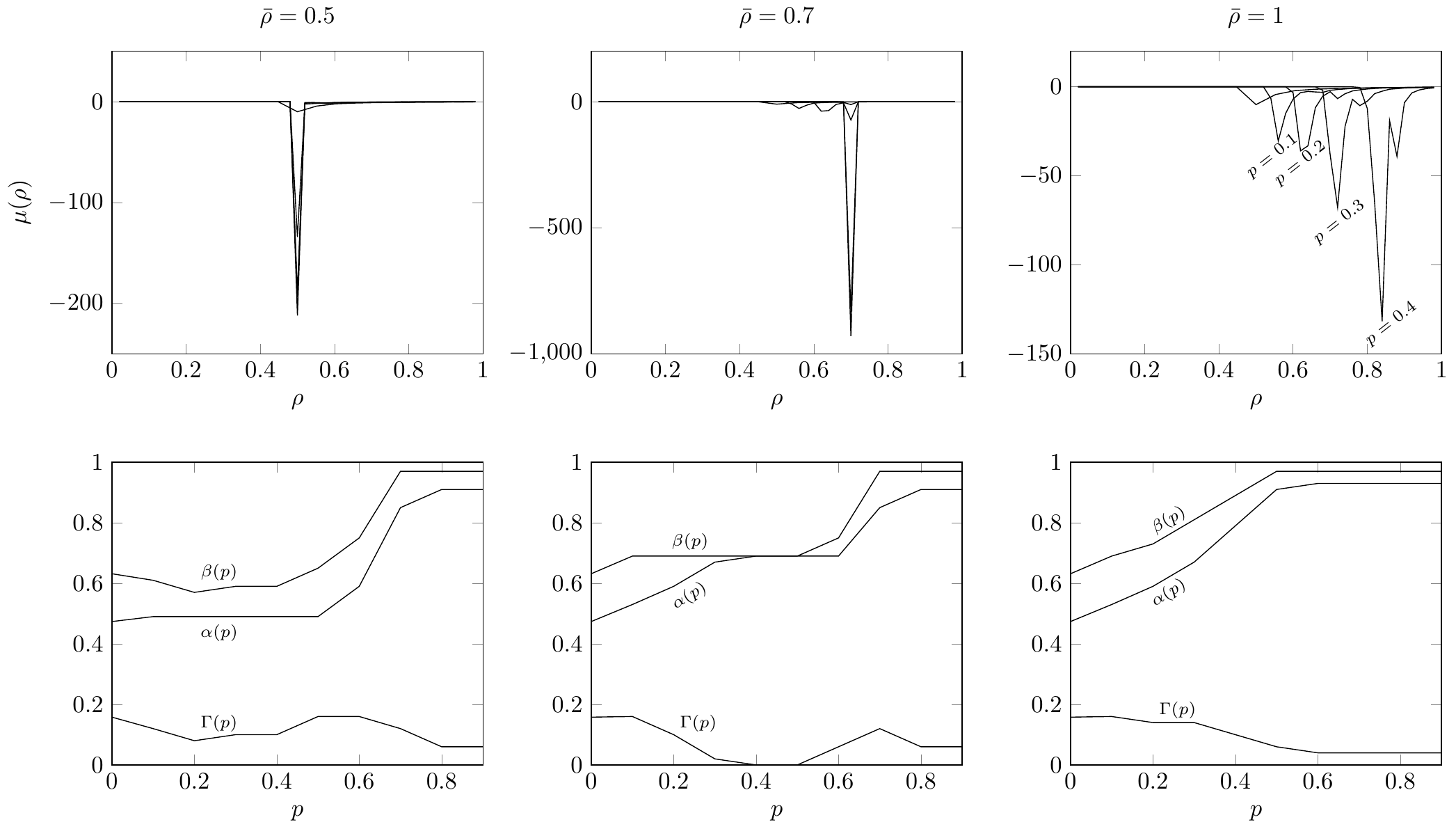}
	\caption{Top row: sign of the diffusion coefficient $\mu(\rho)$ \eqref{eq:newDiffusion} of the single--distribution model~\eqref{eq:singleDistrModel} for autonomous and human--driven vehicles. Three choices of the parameter $\bar{\rho}$ and different penetration rates $p$ are considered. Here $\vmax=1=\rmax=1$, $\Delta v=\frac13$ and $\hat{u}(\rho)=\frac{1}{\rho}\int_{\mathcal{V}} v f(t,v) \d v$. Bottom row: left boundary $\alpha$ and right boundary $\beta$ of the interval of instability, and its amplitude $\Gamma$, as function of the penetration rate $p$.\label{fig:diffusionCA}}
\end{figure}

Secondly, we study the possible appearance of stop and go waves by investigation of the interval of instability $[\alpha,\beta]$ defined in Definition~\ref{def:stabilityBGK}. In the top row of Figure~\ref{fig:diffusionCA} we show the sign of the diffusion coefficient $\mu(\rho)$ \eqref{eq:newDiffusion} of the single--distribution model~\eqref{eq:singleDistrModel} for autonomous and human--driven vehicles. In the bottom row of Figure~\ref{fig:diffusionCA}, in particular, we focus on the analysis of the interval of instability by showing the left boundary $\alpha$ and the right boundary $\beta$ of the regime where $\mu(\rho)<0$ as function of the penetration rate $p$ of the autonomous vehicles. $\Gamma=|\beta-\alpha|$ gives information on the amplitude of the interval of instability as function of $p$. Here, we have considered ten equally spaced values of $p$ in $[0,0.9]$. We notice that the interval of instability moves towards larger density regimes as the penetration rate increases. Moreover, the maximum amplitude of the region is observed for $p=0$, i.e.~when no autonomous vehicles are present on the road. This means that, the regime where the diffusion coefficient is negative moves in very congested traffic regimes as the amount of autonomous vehicles increases.

\section{Conclusion} \label{sec:conclusion}

In this paper we studied the impact of autonomous vehicles on traffic stabilization using a kinetic approach. \revision{In our work autonomous vehicles are endowed with sensors which provide information on the surrounding environment to their drivers, human or computer. The instantaneous and precise information is modeled by deterministic binary interactions involving autonomous cars.} We focused on two indicators of instability. These are the sign of the diffusion coefficient obtained via Chapman--Enskog expansion of the kinetic model and the variability of microscopic speeds at equilibrium. In particular, regimes where the diffusion is negative correspond to a growth of density perturbations, which may be associated to stop and go waves. Instead a high variance of the speeds of vehicles is a source of high risk of collisions.

We analyzed the indicators of instability assuming that a percentage of the traffic density is given by autonomous cars which differ from human--driven cars in the interaction rules\revision{: the latter are characterized by the stochastic behavior of the drivers.} The kinetic model is derived for a single distribution and we analyzed the macroscopic properties of the model. In particular, numerical simulations showed that autonomous cars may help to stabilize traffic flow by reducing variability of microscopic speeds at equilibrium and by damping the effect of instabilities due to stop and go waves. 

\section*{Acknowledgments}

M.H. and G.V. thank the Deutsche Forschungsgemeinschaft (DFG, German Research Foundation) for the financial support through 20021702/GRK2326, 333849990/IRTG-2379, HE5386/15,18-1,19-1,22-1 and under Germany's Excellence Strategy EXC-2023 Internet of Production 390621612. The funding through HIDSS-004 is acknowledged.

G.P. and G.V. acknowledge support from ``National Group for Scientific Computation (GNCS-INDAM)'' and by MUR (Ministry of University and Research) PRIN2017 project number 2017KKJP4X.

\clearpage

\appendix

\section{Nanbu--like algorithm} \label{app}

\begin{algorithm}[h!]
{\normalsize
	\caption{Nanbu algorithm~\cite{Nanbu} for the model~\eqref{eq:singleDistrModel} with interactions~\eqref{eq:HtoHA}--\eqref{eq:AtoA}--\eqref{eq:AtoC} and $\gamma=1$.\label{alg:Nanbu}}
	\begin{algorithmic}[1]
		\STATE Fix $\rho$ being the initial density of vehicles, with $\rho_{\AV}=p\rho$, $\rho_{\HV}=(1-p)\rho$, $p\in[0,1]$;
		\STATE Fix the probability of changing velocity $p(\rho)$ and $\Delta v$;
		\STATE Take $N$ samples of the microscopic velocities $v_j^0$, $j=1,\dots,N$ from the initial density $f(t=0,v)$;
		\FOR{$n=0$ \TO $M$}
		\STATE compute the mean speed $u^n=\frac{1}{N} \sum_{j=1}^N v_j^n$;
		\FOR{$j=1$ \TO $N$}
		\STATE Select three uniformly distributed pseudorandom numbers $r_1$, $r_2$ and $r_3$, and an index $k\in\{1,\dots,N\}$
		\IF{$r_1\leq\rho_\HV$}
		\IF{$r_3\leq P(\rho)$}
		\STATE Compute $v_j^{n+1}=\min\{v_j^n+\Delta v,V_M\}$
		\ELSE
		\IF{$v_j^n\leq v_k^n$}
		\STATE Compute $v_j^{n+1}=v_j^n$
		\ELSE
		\STATE Compute $v_j^{n+1}=v_k^n$
		\ENDIF
		\ENDIF
		\ELSE
		\IF{$r_2\leq\rho_\AV$}
		\STATE Compute $v_j^{n+1}=\min\{v_j^n+\Delta v,\hat{u}(\rho)\}$
		\ELSE
		\IF{$\rho\leq\bar{\rho}$}
		\STATE Compute $v_j^{n+1}=\min\{v_j^n+\Delta v,\hat{u}(\rho)\}$
		\ELSE
		\IF{$v_j^n\leq v_k^n$}
		\STATE Compute $v_j^{n+1}=v_j^n$
		\ELSE
		\STATE Compute $v_j^{n+1}=v_k^n$
		\ENDIF
		\ENDIF
		\ENDIF
		\ENDIF
		\ENDFOR
		\ENDFOR
	\end{algorithmic}
}
\end{algorithm}

\clearpage

\bibliographystyle{plain}
\bibliography{references}

\end{document}